\documentclass[aps,prl,twocolumn,superscriptaddress]{revtex4}%
\usepackage{epsfig,dsfont,amssymb,amsmath,amsthm,amsfonts,amsbsy,mathrsfs}
\usepackage{graphicx}
\usepackage{amsmath}
\usepackage{amssymb}
\usepackage{color}%
\begin{document}

\title{Water on Silicene: Hydrogen Bond Autocatalysis Induced Physisorption-Chemisorption-Dissociation Transition}

\author{Wei Hu}
\affiliation{Hefei National Laboratory for Physical Sciences at
Microscale, Department of Chemical Physics, and Synergetic
Innovation Center of Quantum Information and Quantum Physics,
University of Science and Technology of China, 96 JinZhai Road,
Hefei, Anhui 230026, China} \affiliation{Computational Research
Division, Lawrence Berkeley National Laboratory, 1 Cyclotron Road,
Berkeley, California 94720, United States}

\author{Zhenyu Li}
\affiliation{Hefei National Laboratory for Physical Sciences at
Microscale, Department of Chemical Physics, and Synergetic
Innovation Center of Quantum Information and Quantum Physics,
University of Science and Technology of China, 96 JinZhai Road,
Hefei, Anhui 230026, China}

\author{Jinlong Yang}
\thanks{Corresponding author. E-mail: jlyang@ustc.edu.cn}
\affiliation{Hefei National Laboratory for Physical Sciences at
Microscale, Department of Chemical Physics, and Synergetic
Innovation Center of Quantum Information and Quantum Physics,
University of Science and Technology of China, 96 JinZhai Road,
Hefei, Anhui 230026, China}

\date{\today}

\pacs{ }

\begin{abstract}

ABSTRACT: A single water molecule has nothing special. However,
macroscopic water displays many anomalous properties at the
interface, such as a high surface tension, hydrophobicity and
hydrophillicity. Although the underlying mechanism is still elusive,
hydrogen bond is expected to have played an important role. An
interesting question is if the few-water molecule clusters will be
qualitatively different from a single molecule. Using adsorption
behavior as an example, by carefully choosing two-dimensional
silicene as the substrate material, we demonstrate that water
monomer, dimer and trimer show contrasting properties. The
additional water molecules in dimer and trimer induce a transition
from physisorption to chemisorption then to dissociation on
silicene. Such a hydrogen bond autocatalytic effect is expected to
have a broad application potential in silicene-based water molecule
sensor and metal-free catalyst for water dissociation.

\vspace{3ex}

KEYWORDS: Water adsorption, hydrogen bond, silicene, physisorption,
chemisorption, dissociation and density functional theory

\end{abstract}

\maketitle

\section{Introduction}

Hydrogen bonding is crucial in many molecular and supramolecular
systems, such as water, polymers and proteins. It determines many of
their chemical and physical properties.\cite{AccChemRes_23_120_1990,
AccChemRes_29_441_1996, AccChemRes_38_386_2005} Hydrogen bonding in
water plays an important role in distinguishing it from other
systems with comparable molecular mass. At the same time, water is
also unique among other hydrogen-bonded systems since it has a very
large number of hydrogen bonds per unit mass. Beyond bulk
properties, the structure of water at the interface with other
materials can induce many anomalous properties, such as a high
surface tension of water, hydrophobicity and
hydrophillicity.\cite{ChemRev_106_1478_2006} In fact, water at the
interface plays a critical role in wide aspects ranging from daily
life to science and technology.

In recent years, two-dimensional (2D) ultrathin
materials,\cite{PNAS_102_10451_2005, AdvMater_24_210_2012,
NatureNanotechnol_7_699_2012, ChemRev_113_3766_2013} such as
graphene,\cite{Scinece_306_666_2004, NatureMater_6_183_2007,
RMP_81_109_2009} silicene,\cite{PRB_50_14916_1994,
PRB_76_075131_2007, PRL_102_236804_2009, SSR_67_1_2012}
germanene,\cite{ACSNano_7_4414_2013, AdvMater_26_4820_2014,
NJP_16_095002_2014} phosphorene,\cite{NatureNanotech_9_372_2014,
ACSNano_8_4033_2014, NatureCommun_5_4475_2014} hexagonal boron
nitride (h-BN),\cite{NatureMater_3_404_2004, NanoLett_10_3209_2010,
NanoLett_10_4134_2010} and molybdenum disulphide
(MoS$_2$),\cite{PRL_105_136805_2010, NatureNanotechnol_6_147_2011,
ACSNano_6_74_2012} have received considerable interest owing to
their remarkable properties and wide applications. In particular,
graphene,\cite{Scinece_306_666_2004, NatureMater_6_183_2007,
RMP_81_109_2009} a 2D sp$^2$-hybridized carbon sheet, is known to
have remarkable electronic properties, such as a high carrier
mobility. The water-graphene interface also been widely studied
experimentally\cite{Nature_6_652_2007, small_6_2535_2010,
ACSNano_4_1321_2010, NanoLett_12_1459_2012, JACS_134_5662_2012} and
theoretically,\cite{PRL_95_136105_2005, PRB_78_075442_2008,
PRB_77_125416_2008, PRB_79_235440_2009, PRB_79_113409_2009,
PRB_85_085425_2012} for the purpose to realize charge
doping,\cite{NanoLett_12_1459_2012} charge
transfer\cite{PRB_78_075442_2008} and band-gap
opening\cite{PRB_77_125416_2008} for graphene based field effect
transistors, molecule sensor,\cite{Nature_6_652_2007} metal-free
catalyst for oxygen reduction reaction\cite{ACSNano_4_1321_2010} and
water dissociation.\cite{PRL_95_136105_2005} 2D low-cost graphene
has more large contact area for water adsorption than traditional 3D
noble metals and metal oxides,\cite{ChemRev_106_1478_2006} such as
platinum\cite{JACS_123_4235_2001, PRL_89_276102_2002,
PRL_90_216102_2003} and TiO$_2$,\cite{PRL_87_266103_2001,
PRL_87_266104_2001, Science_308_1154_2005} as the catalysts for
water chemisorption and dissociation. However, water molecules are
physically adsorbed on graphene with very small adsorption energies
via weak van der Waals interactions\cite{PRB_77_125416_2008} and
graphene is strongly hydrophobic,\cite{PRB_79_235440_2009}
preventing immediate practical applications on graphene based
sensitive molecule sensor and efficient catalyst if without
introducing dopants or defects.\cite{ACSNano_4_1321_2010,
JACS_134_5662_2012, PRL_95_136105_2005, PRB_79_113409_2009}

Silicene, analog to graphene but with buckled honeycomb
structures,\cite{PRB_50_14916_1994, PRB_76_075131_2007,
PRL_102_236804_2009} has also attracted increasing attentions for
its excellent properties\cite{SSR_67_1_2012} similar to graphene,
such as high carrier mobility,\cite{JCP_139_154704_2013} electric
response,\cite{PRB_85_075423_2012}
ferromagnetism,\cite{PCCP_14_3031_2012} quantum Hall
effect,\cite{PRL_107_076802_2011} giant
magnetoresistance\cite{Nanoscale_4_3111_2012} and
superconductivity.\cite{APL_102_081602_2013} Silicene has been
widely experimentally fabricated on Ag\cite{NanoLett_12_3507_2012,
PRL_108_155501_2012, PRL_109_056804_2012, PRL_110_085504_2013} and
Ir\cite{NanoLett_13_685_2013} substrates. Due to its buckled
honeycomb structures, silicene exhibits a much higher chemical
reactivity than graphene, showing much stronger adsorption of
atoms\cite{PRB_86_075440_2012, PRB_87_085444_2013,
PRB_87_085423_2013, NanoLett_13_2258_2013, JAP_114_124309_2013,
JPCC_117_483_2013, PRL_111_145502_2013} and
molecules\cite{JPCC_117_26305_2013, PCCP_15_5753_2013,
Nanoscale_5_9062_2013, PCCP_16_6957_2014, CMS_87_218_2014} than
graphene, with great potential applications on new silicene based
nanoelectronic devices,\cite{SSR_67_1_2012} Li-ion storage
batteries,\cite{NanoLett_13_2258_2013} hydrogen
storage,\cite{JAP_114_124309_2013} catalyst,\cite{JPCC_117_483_2013}
thin-film solar cell absorbers,\cite{PRL_111_145502_2013}
hydrogen\cite{PCCP_15_5753_2013} and
helium\cite{Nanoscale_5_9062_2013} separation membrane, molecule
sensor and detection,\cite{PCCP_16_6957_2014, CMS_87_218_2014}
superior to graphene. However, water monomer is still proved to be
physically adsorbed on silicene via van der Waals
interactions,\cite{JPCC_117_26305_2013, PCCP_15_5753_2013} similar
to graphene.\cite{PRB_78_075442_2008, PRB_77_125416_2008,
PRB_79_235440_2009}

In the present work, by using the first-principles density
functional theory calculations and ab-initio molecular dynamics
simulations, we find that hydrogen bonding in water dimer and trimer
can induce autocatalytic chemisorption and dissociation of water
molecules on silicene. Furthermore, the interaction between water
molecules and silicene increases as the number of hydrogen bonding
in water molecules. We also find that silicene is hydrophilic
different from graphene.

\section{Theoretical Models and Methods}

The lattice parameter of silicene calculated to setup unit cell is
3.866 {\AA}, agreeing well with previous theoretical
works.\cite{PCCP_14_3031_2012} In order to simulate the infinite
planar monolayer, a 4 $\times$ 4 supercell of silicene containing 32
silicon atoms is adopted. Different adsorption sites (valley,
bridge, hollow and top) of water molecules on silicene are
considered as shown in Figure~\ref{fig:1H2O}. The vacuum space in
the $Z$ direction is about 20 {\AA} to separate the interactions
between neighboring slabs.

We use the first-principles density functional theory (DFT)
calculations implemented in the VASP
package.\cite{PRB_47_558_1993_VASP} We choose the generalized
gradient approximation of Perdew, Burke, and Ernzerhof
(GGA-PBE)\cite{PRL_77_3865_1996_PBE} as the exchange-correlation
functional, and adopt the semi-empirical van der Waals dispersion
correction proposed by Grimme (DFT-D2)\cite{JCC_27_1787_2006_Grimme}
to describe the weak van der Waals interactions of layered 2D
materials and molecular adsorption on
surfaces.\cite{JPCC_111_11199_2007, PCCP_10_2722_2008,
NanoLett_11_5274_2011, PRB_83_245429_2011, PRB_85_125415_2012,
PRB_85_235448_2012, JPCL_4_2158_2013, JCP_138_124706_2013} We check
the adsorption structures of water molecules on silicone with
non-empirical van der Waals density functional (vdW-DF) scheme
proposed by Dion et al.\cite{PRL_92_246401_2004_Dion} and obtain
similar results. Because the GGA-PBE method trends to underestimate
the bandgap of semiconductors, the screened hybrid HSE06
functional\cite{JCP_124_219906_2006_HSE06} is also used to compute
the electronic band structures. We set the energy cutoff to be 500
eV. The surface Brillouin zone is sampled with a 3 $\times$ 3
regular mesh and 120 (GGA-PBE) or 60 (HSE06) k points are used for
calculating the small band gaps at the Dirac points of silicene. All
the geometry structures are fully relaxed by using the conjugate
gradient (CG) algorithm until total energy and atomic forces are
converged to 10$^{-5}$ eV and 0.01 eV/{\AA}, respectively. Charge
transfer is obtained based on Bader
analysis.\cite{CMS_36_354_2006_Bader} Ab-initio molecular dynamics
(AIMD) simulations are performed in a canonical ensemble with a 4
$\times$ 4 supercell of silicene containing 32 silicon atoms and
four layers of water molecules containing 144 oxygen atoms and 288
hydrogen atoms (1.0 g/cm$^3$). The energy cutoff is set to be 350
eV. The simulations are performed for about 13.0 ps with a time step
of 1.0 fs at the temperature of 300 K controlled by a Nose-Hoover
thermostat.\cite{JCP_81_511_1984_Nose, PRA_31_1695_1985_Hoover}

In order to evaluate the stability of water molecules on silicene,
the adsorption energy is defined as $E_{a}$ =
$E$((H$_2$O)$_n$/Silicene) - $E$((H$_2$O)$_n$) - $E$(Silicene),
where, $E$((H$_2$O)$_n$/Silicene), $E$((H$_2$O)$_n$) and
$E$(Silicene) represent the total energy of water molecules
adsorption on silicene, water molecule clusters and pristine
silicene, respectively. For water molecule clusters ($n$ = 2 and
3.), the binding energy of hydrogen bond is defined as $E_{b}$ =
$E$((H$_2$O)$_n$) - $n$$E$(H$_2$O), where, $E$((H$_2$O)$_n$) and
$E$(H$_2$O) represent the total energy of water molecule clusters
and single water molecule, respectively. As an benchmark, DFT-D2
calculations give a good bilayer distance of 3.25 {\AA} and binding
energy of -25 $meV$ per carbon atom for bilayer graphene, fully
agreeing with previous experimental\cite{PR_100_544_1955,
PRB_69_155406_2004} and theoretical\cite{PRB_85_205402_2012,
JCP_138_054701_2013} studies. Furthermore, DFT-D2 calculations also
give accurate binding energy of -0.26 $eV$ in water molecule
dimer.\cite{JCTC_8_281_2012}

\section{Results and Discussion}

\begin{figure}[htbp]
\begin{center}
\includegraphics[width=0.5\textwidth]{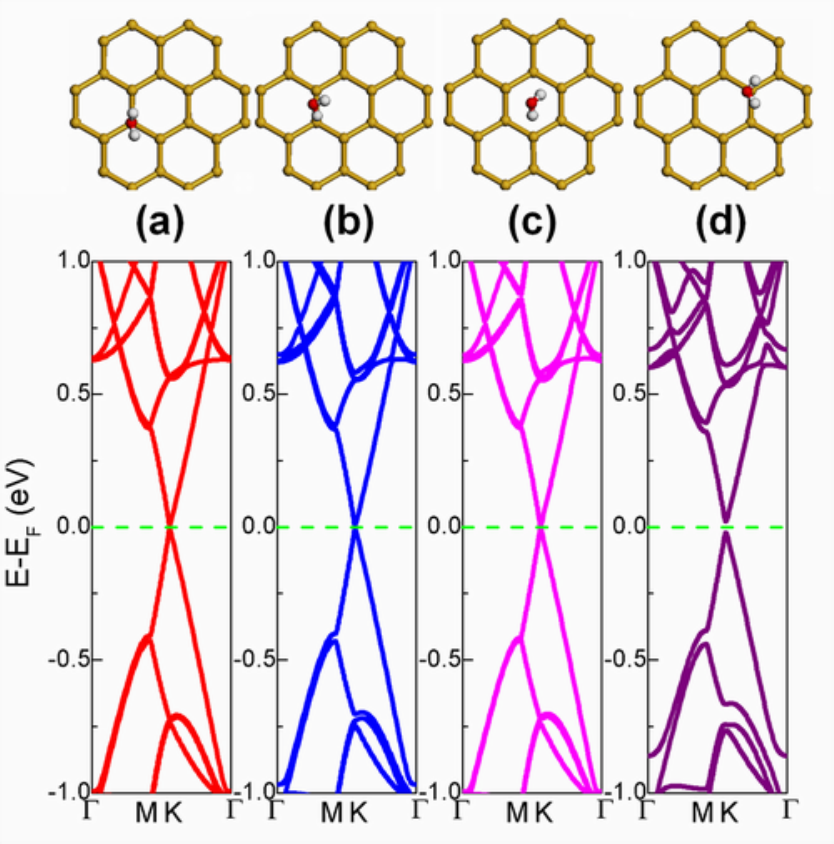}
\end{center}
\caption{(Color online) Geometric and electronic structures (HSE06)
of water monomer physisorbed on silicene. Different adsorption sites
are considered, (a) valley, (b) bridge, (c) hollow and (d) top. The
yellow, red and white balls denote silicon, oxygen and hydrogen
atoms, respectively. The Fermi level is set to zero and marked by
green dotted lines.} \label{fig:1H2O}
\end{figure}

We first check the adsorption of water monomer on silicene.
Geometric and electronic structures for different adsorption sites
are shown in Figure~\ref{fig:1H2O} and the corresponding adsorption
properties are listed in Table 1. Water monomer is physically
adsorbed on silicene via van der Waals
interactions,\cite{JPCC_117_26305_2013, PCCP_15_5753_2013} with
small adsorption energies (-0.13 $\sim$ -0.17 eV) and large
adsorption distances (2.49 $\sim$ 2.92 {\AA}). The top site is the
most stable adsorption site. Water adsorption on silicene acts as an
electron acceptor (0.08 $\sim$ 0.23 $e$), similar to
graphene.\cite{PRB_78_075442_2008, PRB_77_125416_2008,
PRB_79_235440_2009}

\begin{table}
\caption{Adsorption properties of water molecules adsorption on
silicene, including adsorption type (Phys/Chem/Diss), adsorption
energy $E_{a}$ (eV), adsorption distance (or Si-O bond length) $d$
({\AA}), electron transfer from silicene to water molecules $\rho$
($e$) and band gap $E_g$ (meV) opened at the Dirac point for water
molecules on silicene.}
\begin{tabular}{ccccccc} \\ \hline \hline
H$_2$O/Silicene                 &  Phase & $E_{a}$ &  $d$  & $\rho$ & $E_g$ \ \\
\hline
(H$_2$O)$_1$/Silicene (Valley)  &  Phys  &  -0.13  &  2.87  &  0.21  &  11   \ \\
(H$_2$O)$_1$/Silicene (Bridge)  &  Phys  &  -0.15  &  2.92  &  0.14  &  23  \ \\
(H$_2$O)$_1$/Silicene (Hollow)  &  Phys  &  -0.16  &  2.89  &  0.23  &  13  \ \\
(H$_2$O)$_1$/Silicene (Top)     &  Phys  &  -0.17  &  2.49  &  0.08  &  66  \ \\
(H$_2$O)$_2$/Silicene (Top)     &  Chem  &  -0.67  &  1.93  &  0.23  &  121  \ \\
(H$_2$O)$_3$/Silicene (Top)     &  Chem  &  -1.19  &  1.81  &  0.44  &  128  \ \\
(H$_2$O)$_3$/Silicene (Top)     &  Diss  &  -1.09  &  1.77  &  0.50  &  152  \ \\
(H$_2$O)$_1$/Silicene/Ag (Top)  &  Phys  &  -0.28  &  2.33  &  0.21  &  -  \ \\
(H$_2$O)$_2$/Silicene/Ag (Top)  &  Chem  &  -0.93  &  1.90  &  0.51  &  -  \ \\
(H$_2$O)$_3$/Silicene/Ag (Top)  &  Chem  &  -1.64  &  1.79  &  0.85  &  -  \ \\
(H$_2$O)$_3$/Silicene/Ag (Top)  &  Diss  &  -1.58  &  1.75  &  0.73  &  -  \ \\
\hline \hline
\end{tabular}
\end{table}

Silicene's linear Dirac-like dispersion relation $E(k) =
{\pm}{\hbar}{\nu}_F|k|$ around the Fermi level is slightly affected
by adsorption of water monomer. The Fermi velocity ${\nu}_F
=(1/{\hbar})({\eth}E/{\eth}k)$ is about $10^5$ m/s for water
adsorbed silicene, only slightly lower than that of the freestanding
silicone. Therefore, silicene's high carrier mobility can be
preserved for water molecules adsorption. Upon water molecule
adsorption, a small band gap is opened at the Dirac point of
silicene. For adsorption at the valley, bridge, and hollow sites,
the energy gap is 11, 23 and 13 meV, respectively. They are
significantly smaller than thermal fluctuation (about 25 meV) at
room temperature, similar to the graphene case. A wider band gap of
66 meV is opened at the Dirac point if water monomer is adsorbed at
the top site.

\begin{figure}[htbp]
\begin{center}
\includegraphics[width=0.5\textwidth]{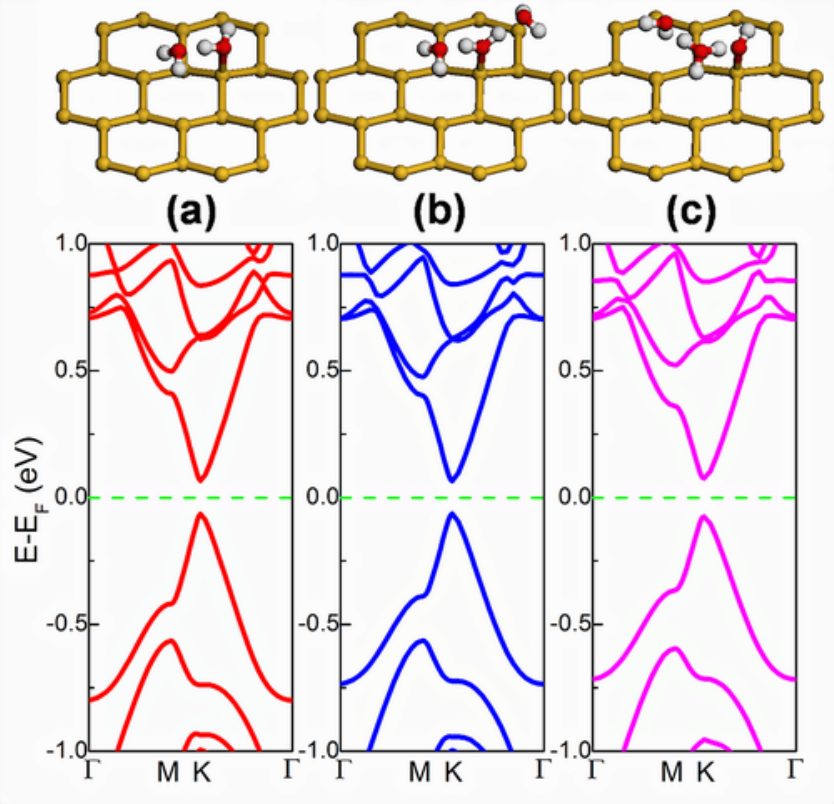}
\end{center}
\caption{(Color online) Geometric and electronic structures (HSE06)
of water dimer and trimer adsorption on silicene, including the
chemisorption of (a) (H2O)$_2$/Silicene and (b) (H2O)$_3$/Silicene,
and (c) the dissociation of (H2O)$_3$/Silicene. The yellow, red and
white balls denote silicon, oxygen and hydrogen atoms, respectively.
The Fermi level is set to zero and marked by green dotted lines.}
\label{fig:2H2O}
\end{figure}

As shown in Figure~\ref{fig:2H2O}, interestingly, water molecule
dimer and trimer are chemically adsorbed and even dissociated on
silicene with large adsorption energies (-0.67, -1.19 and -1.09 eV)
via strong covalent Si-O bonds (1.93, 1.81 and 1.77 {\AA}) between
an oxygen atom (O$_a$) in water and a top silicon atom. Notice that
the adsorption energies predicted here are much larger than the
binding energy of hydrogen bond formation in water clusters (only
-0.26 eV per hydrogen bond). Therefore, this result suggests that
silicene is hydrophilic in contrast to
graphene.\cite{PRB_79_235440_2009}

Compared to single molecule case, more sizable band gaps (121, 128
and 152 meV) are opened at the Dirac point of silicene for water
dimer and trimer chemisorption on silicene and water trimer
dissociation on silicene, which are significantly larger than
thermal fluctuation (25 meV) at room temperature. Furthermore, these
gap values also depend sensitively on the adsorbate concentration.
Therefore, water cluster adsorbed silicene systems have sizable and
tunable band gaps with potential application on silicene based water
molecule sensor and field-effect transistors.

\begin{figure}[htbp]
\begin{center}
\includegraphics[width=0.5\textwidth]{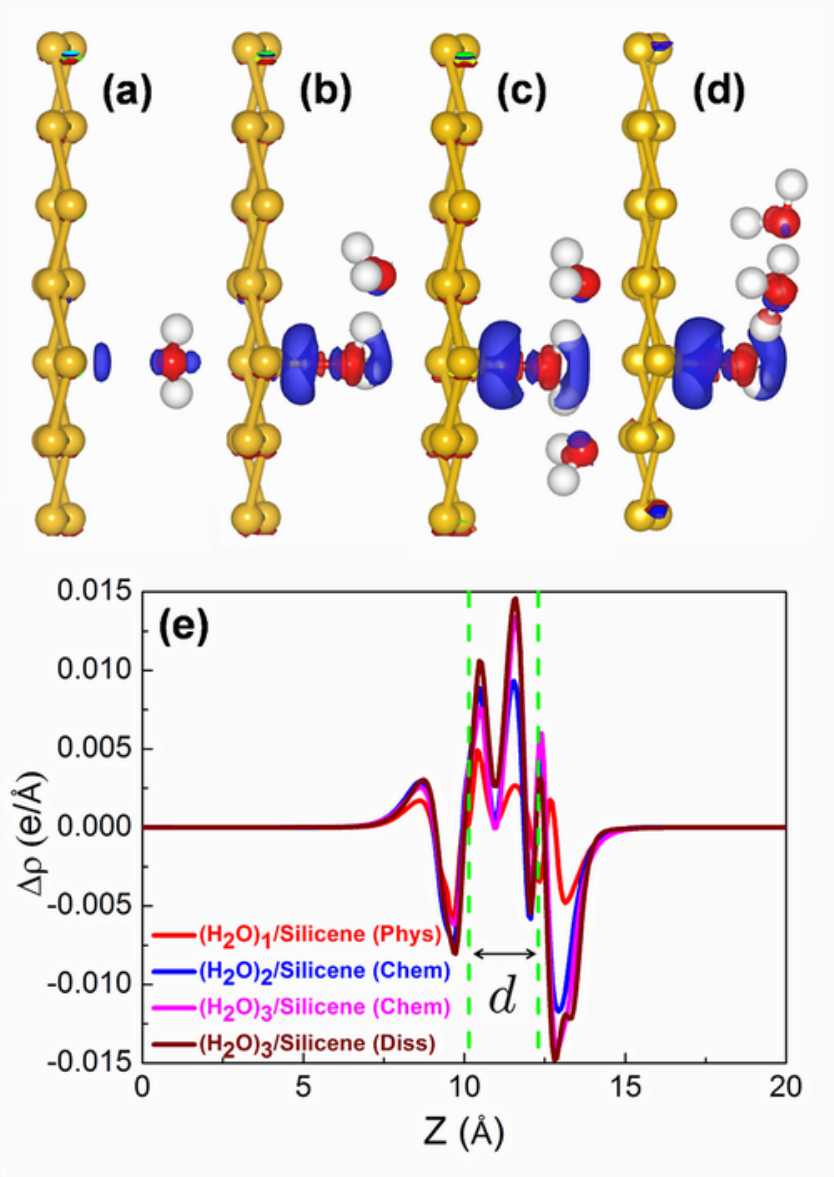}
\end{center}
\caption{(Color online) Differential charge density of water
molecules adsorption on silicene (0.04 $e$/{\AA}$^3$), (a)
physisorption of (H$_2$O)$_1$/Silicene, (b) chemisorption of
(H$_2$O)$_2$/Silicene, (c) chemisorption of (H$_2$O)$_3$/Silicene
and (d) dissociation of (H$_2$O)$_3$/Silicene. Red (positive) and
blue (negative) regions indicate electron increase and decrease,
respectively. (e) XY-averaged differential charge density. The
water-silicene interface is marked by green dotted lines.}
\label{fig:Charge}
\end{figure}

Figure~\ref{fig:Charge} shows the differential charge density of
water molecules adsorption on silicene ($\triangle$$\rho$ =
$\rho$((H$_2$O)$_n$/Silicene) - $\rho$((H$_2$O)$_n$) -
$\rho$(Silicene)) and the corresponding XY-averaged differential
charge density. We find that more electrons (0.23, 0.44 and 0.50
$e$) transfer from silicene to water in dimer and trimer adsorption,
and trimer dissociation, compared to that (0.08 $e$) in the single
water molecule case. Therefore, the interaction between water
molecules and silicene increases as the number of hydrogen bonds
formation in water molecules increases.

\begin{figure}[htbp]
\begin{center}
\includegraphics[width=0.5\textwidth]{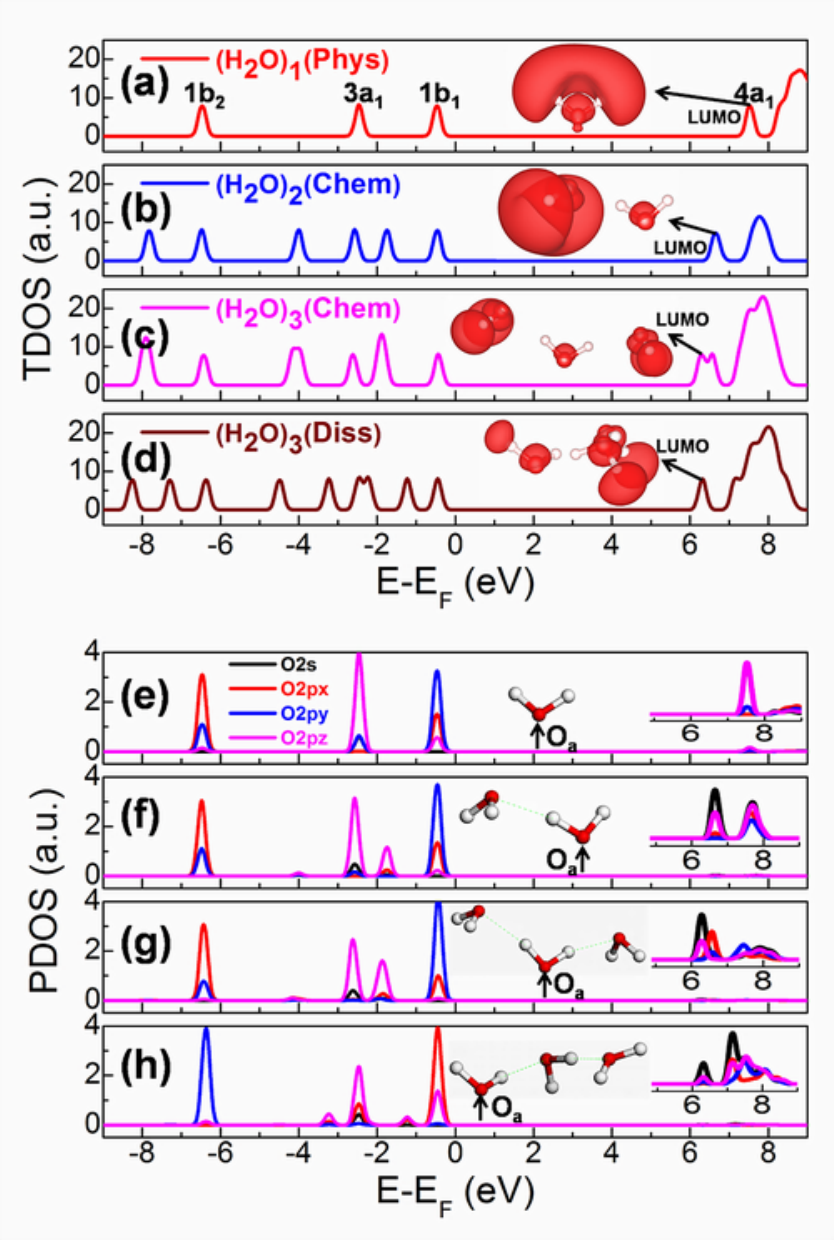}
\end{center}
\caption{(Color online) Electronic structures (HSE06) of water
clusters ((H$_2$O)$_1$, (H$_2$O)$_2$ and (H$_2$O)$_3$). Total
density of states (TDOS) of water molecules, (a) (H$_2$O)$_1$(Phys),
(b) (H$_2$O)$_2$(Chem), (c) (H$_2$O)$_3$(Chem) and (d)
(H$_2$O)$_3$(Diss). LUMO states of water molecules are shown in the
inset. Partial density of states (PDOS) of the adsorption oxygen
atom (O$_a$) of water molecules at the top of silicene, (e)
(H$_2$O)$_1$(Phys), (f) (H$_2$O)$_2$(Chem), (g) (H$_2$O)$_3$(Chem)
and (h) (H$_2$O)$_3$(Diss). Magnified LUMO is shown in the inset.
The Fermi levels of water molecules are set to zero.}
\label{fig:H2O}
\end{figure}

Figure~\ref{fig:H2O} shows total density of states (DOS) of water
molecules ((H$_2$O)$_1$, (H$_2$O)$_2$ and (H$_2$O)$_3$) and the
corresponding partial density of states (PDOS) of the oxygen atom
(O$_a$) which binds to silicene upon adsorption. The frontier
orbitals 1b$_1$ (highest occupied molecular orbital (HOMO) and
lowest unoccupied molecular orbital 4a$_1$ (LUMO)) of H$_2$O are far
away from the Dirac point of graphene and
silicene.\cite{PCCP_16_6957_2014} Thus, water monomer has low
reactivity to both graphene and silicene. But for water dimer and
trimer, the LUMO gets lower and becomes closer to the Fermi level as
more hydrogen bonds are formed in water clusters (7.5, 6.6, 6.3 and
6.3 eV respectively for water monomer physisorption, water dimer
chemisorption, trimer chemisorption and dissociation on silicene).
Especially for for the LUMO states contributed by the O$_a$ atom, we
find significant energy shift of LUMO in water molecules makes
electron transfer from silicene to water easier. Notice that the
O$_a$ atom in the structure of water trimer for dissociation on
silicene shows more localized LUMO states compared to that in water
trimer for chemisorption on silicene. Therefore, the activated O$_a$
atom can form a Si-O bond upon adsorption, showing an autocatalytic
behavior for chemisorption and dissociation compared to single water
molecule adsorption. Hydrogen bonding in water thus plays an
important role in its adsorption at the H$_2$O/Silicene interface.

We notice that water cluster is physically adsorbed on graphene,
which can be understood from the electronic structure difference
between graphene and silicene.\cite{PCCP_15_5753_2013} Although they
have similar work functions, the reactive $p_z$ state is closer to
the Fermi level in silicene compared to the graphene case. We also
check other elemental 2D materials with higher chemical reactivity
than graphene, such as germanene and phosphorene. Though oxygen
molecules are easily chemically adsorbed and dissociated on
silicene,\cite{JPCC_117_26305_2013}
germanene\cite{PCCP_16_22495_2014} and
phosphorene\cite{PRL_114_046801_2015} at room temperature, we find
that water molecules are only chemically adsorbed and dissociated on
silicene. Therefore, silicene should be a special 2D substrate
material for adsorbing water molecules.

\begin{figure}[htbp]
\begin{center}
\includegraphics[width=0.5\textwidth]{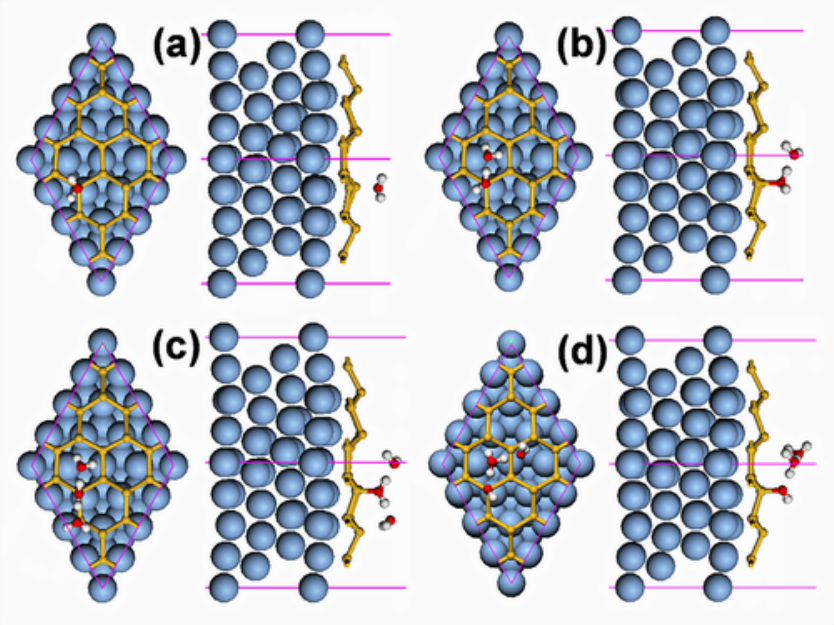}
\end{center}
\caption{(Color online) Geometric structures of water molecules
adsorption on silicene supported by the Ag(111) surface, including
(a) physisorption of (H$_2$O)$_1$/Silicene/Ag, (b) chemisorption of
(H$_2$O)$_2$/Silicene/Ag, (c) chemisorption of
(H$_2$O)$_3$/Silicene/Ag and (d) dissociation of
(H$_2$O)$_3$/Silicene/Ag. The yellow, red, white and blue balls
denote silicon, oxygen, hydrogen and silver atoms, respectively.}
\label{fig:H2OAg}
\end{figure}

To date, it is still a challenge to obtain freestanding silicene. To
confirm that the novel adsorption behavior is universal, we also
consider silicene on a typical substrate, the Ag(111)
surface.\cite{NanoLett_12_3507_2012, PRL_108_155501_2012,
PRL_109_056804_2012, PRL_110_085504_2013} A 3 $\times$ 3 silicene
supercell is used to match a 4 $\times$ 4 supercell of the Ag(111)
surface, which gives a negligible lattice mismatch
(2\%).\cite{PRL_108_155501_2012} As shown in Figure~\ref{fig:H2OAg},
water monomer is still physically adsorbed on Ag-supported silicene
($d_{Si-O}$ = 2.33 {\AA} and Ea = -0.28 eV), and water dimer
($d_{Si-O}$ = 1.90 {\AA} and Ea = -0.93 eV) and trimer ($d_{Si-O}$ =
1.79 {\AA} and Ea = -1.64 eV) are chemisorbed on Ag-supported
silicene. Furthermore, more electrons (0.21, 0.51, 0.85 and 0.73
$e$) transfer from Ag-supported silicene to water compared to that
(0.08, 0.23, 0.44 and 0.50 $e$) of water molecules adsorbed on
free-standing silicene. Therefore, the interaction between water
molecules and silicene is strengthened when an Ag substrate is
presented, similar to water molecules adsorption on metal-supported
grapheme.\cite{PRB_85_085425_2012} Although the substrates do have
an effect on water adsorption, the autocatalytic behavior we
observed is quite robust.

\begin{figure}[htbp]
\begin{center}
\includegraphics[width=0.5\textwidth]{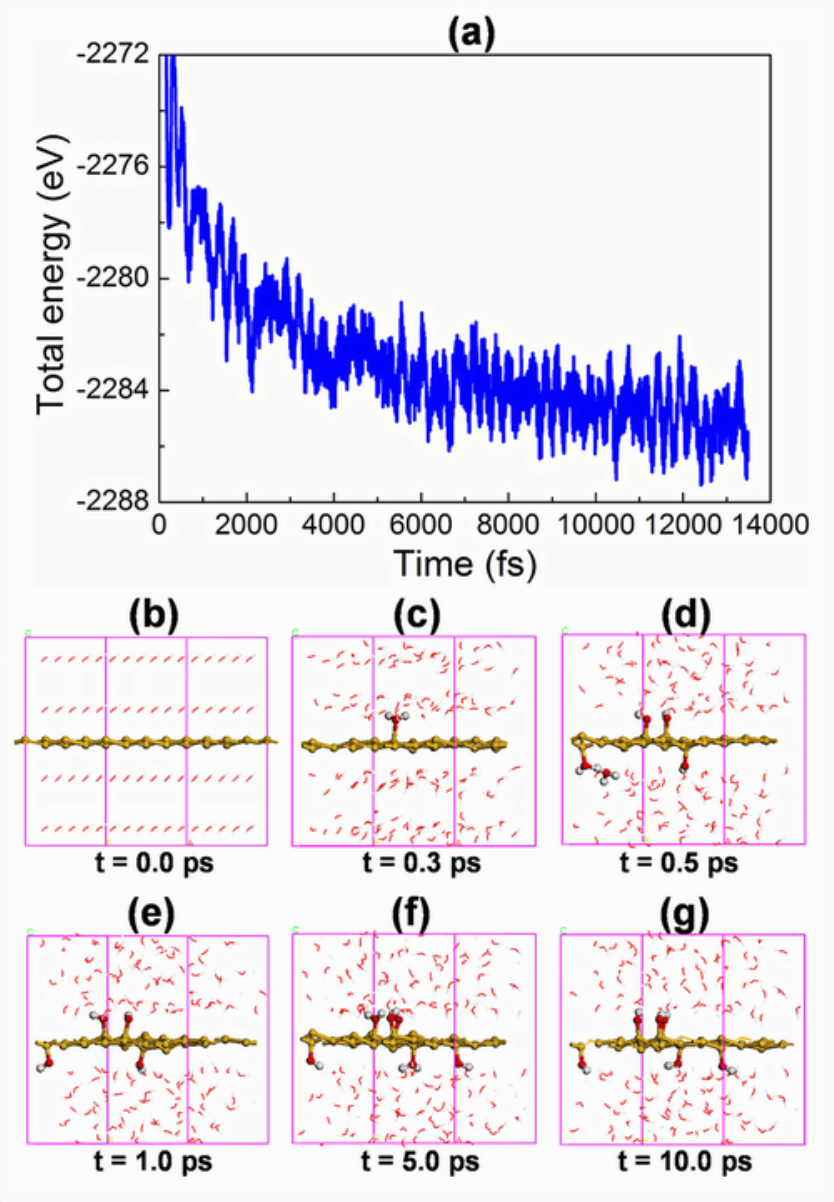}
\end{center}
\caption{(Color online) AIMD simulations of water molecules
adsorption on silicene. (a) AIMD fluctuations of total energy during
13.0 ps at 300 K. Six snapshots of water molecules adsorption on
silicene are shown in the inset, (b) t = 0.0 ps, (c) t = 0.3 ps, (d)
t = 0.5 ps,  (e) t = 1.0 ps, (f) t = 5.0 ps and (g) t = 10.0 ps.
Water molecules chemically adsorbed and dissociated on silicene are
highlighted in the insert.} \label{fig:MD}
\end{figure}

To further confirm the physical picture obtained, AIMD simulations
are also performed at room temperature (300 K). As shown in
Figure~\ref{fig:MD}, in the initial configuration (t = 0.0 ps),
water molecules are set to be physically adsorbed on silicene. At t
= 0.3 ps, we find that one water molecule is chemically adsorbed on
silicene. At t =0.5 ps, more water molecules are chemically adsorbed
and some of them are even dissociated on silicene at room
temperature. At t =1.0 ps, most of water molecules are dissociated
on silicene. After t =10.0 ps, more and more water molecules are
chemically adsorbed and dissociated on silicene at room temperature.
Therefore, the chemisorption of water molecules on silicene and the
hydrophobicity of silicene provide potential applications on
silicene based water molecule sensor and metal-free catalyst for
oxygen reduction reaction and water dissociation without introducing
dopants or defects.\cite{ACSNano_4_1321_2010, JACS_134_5662_2012,
PRL_95_136105_2005, PRB_79_113409_2009}

\section{Summary and Conclusions}

In summary, we have explored the interaction between water molecules
and silicene via density functional theory calculations and
ab-initio molecular dynamics simulations. We find that water monomer
interacts weakly with silicene via a van der Waals interaction. But,
due to the hydrogen bond induced charge transfer, water clusters
(dimer and trimer) are chemically adsorbed and dissociated on
silicene via strong covalent Si-O bonds. Charge transfer occurs from
silicene to water molecules. Our calculations show that silicene is
hydrophilic different from other widely studied two-dimensional
materials, and the chemisorption and dissociation of water molecule
clusters on silicene have immediate applications as water molecule
sensor and metal-free catalyst for oxygen reduction reaction and
water dissociation without introducing
dopants or defects, superior to graphene. \\

\section{ACKNOWLEDGMENTS}

This work is partially supported by the National Key Basic Research
Program (2011CB921404), by NSFC (11404109, 21121003, 91021004,
21233007, 21222304), by CAS (XDB01020300). This work is also
partially supported by the Scientific Discovery through Advanced
Computing (SciDAC) Program funded by U.S. Department of Energy,
Office of Science, Advanced Scientific Computing Research and Basic
Energy Sciences (W. H.). We thank the National Energy Research
Scientific Computing (NERSC) center, and the USTCSCC, SC-CAS,
Tianjin, and Shanghai Supercomputer Centers for the computational
resources.

\vspace{3ex}

\footnotesize{
\bibliography{achemso}
}

\vspace{3ex}


\end{document}